\documentclass[aps,prl,reprint,nofootinbib]{revtex4-1}
\usepackage{graphicx}
\usepackage[latin1]{inputenc}
\usepackage{epsfig}
\usepackage{amsmath}
\usepackage{amsfonts}
\usepackage{amssymb}
\usepackage{perpage} 
\usepackage{slashed}
\MakePerPage{footnote}
\def\lsim{\:\raisebox{-0.5ex}{$\stackrel{\textstyle<}{\sim}$}\:}

\begin{document}

\title {A Holographic Twin Higgs Model}
\author{Michael Geller}
\email{mic.geller@gmail.com}
\affiliation{Physics Department, Technion-Institute of Technology, Haifa 32000, Israel}
\author{Ofri Telem}
\email{t10ofrit@gmail.com}
\affiliation{Physics Department, Technion-Institute of Technology, Haifa 32000, Israel}

\begin{abstract}
We present a UV completion of the twin Higgs idea in the framework of holographic composite Higgs. The SM contribution to the Higgs potential is effectively cut off by the SM-singlet mirror partners at the sigma-model scale $f$, naturally allowing for $m_{KK}$ beyond the LHC reach.
The bulk symmetry is $\small SU(7)\times SO(8)$, broken on the IR brane into $\small SU(7)\times SO(7)$ and on the UV brane into $\small (SU(3)\times SU(2)\times U(1))^{SM} \times (SU(3)\times SU(2)\times U(1))^{mirror}\times Z_2$. The field content on the UV brane is the SM, extended by a sector transforming under the mirror gauge group, with the $Z_2$ exchanging the two sectors. An additional $Z_2$ breaking term is generated holographically to reproduce the Higgs mass and VEV, with a mild ${\cal O}(10\%)$ tuning. This model has no trace at the LHC, but can by probed by precision Higgs measurements at future lepton colliders, and by direct searches for KK excitations at a 100 TeV collider.
\end{abstract}


\maketitle

\section{Introduction}

The measured mass of the Higgs \cite{Higgs_Discovery} and the absence of any new discovery in the first run of the LHC, severely constrain the theory space of allowed BSM models. Specifically, to survive the severe bounds, a model has to account for a Higgs mass much lower than the new states which restore naturalness. The composite Higgs framework \cite{Composite} is a class of models, which address this problem by realizing the Higgs as a pseudo Nambu-Goldstone boson (PNGB) of a broken global symmetry. Loop corrections to the Higgs mass are controlled by collective breaking mechanisms - for example in the ``little Higgs"\cite{Little_Higgs} scenario and in holographic Higgs models \cite{MCHM_ADP,MCHM_ACP,MCHM_CNP}. In the latter, the composite Higgs spectrum is calculated using 5d holography \cite{holographic_fermions,Holography}, and the collective breaking is due to 5d locality.

In composite Higgs models the Higgs quadratic is generated mostly via top loops, cut off at the scale of fermionic excitations $m_{\psi}$ where the top partners restore naturalness. It scales naively as $\mu^2 \sim \frac{3}{8\pi^2}y_t^2 m^2_{\psi}$, where $y_t$ is the top Yukawa. Hence, for values of $m_{\psi}$ larger than 1 TeV, the Higgs potential has to be tuned. Direct searches for vector-like top partners \cite{VLQ_searches} put a lower limit on $m_{\psi}$ and therefore on the amount of tuning required to get the correct Higgs potential \cite{tune-MCHM,tune-Panico}. The future runs of the LHC will be able to probe top partners up to $\sim 2$ TeV \cite{VLQ_whitepaper}, and the lack of any discovery would be translated to a percent level tuning \cite{tune-Panico}.

A counterexample to this link between LHC non-discovery and tuning is the possibility that the top partners are light and colorless - and cannot be detected at the LHC. The ``Twin Higgs" model \cite{Twin_Higgs,Twin_Nomura,Orbifold_Higgs} is a realization of this idea in which the top partners are singlets of the entire SM gauge group. In this model the gauge symmetry is extended to $(SU(3)\times SU(2) \times U(1))^{SM} \times (SU(3)\times SU(2) \times U(1))^{m}$ and the Higgs is a PNGB of the breaking of a global $SU(4)/SU(3)$. Additionally, a $Z_2$ symmetry is postulated, exchanging SM particles with their mirror partners, charged only under the mirror gauge group. The global symmetry breaking pattern in this model ensures that the SM contribution to the Higgs potential is cancelled by the contribution of the mirror partners. The effective cutoff in the loops is the mass of the top mirror partner given by $y_tf$, where $f$ is the sigma-model scale. The quadratic term scales as \cite{Twin_Higgs}:
\begin{equation}
\mu^2 \sim  f^2 \frac{3}{8\pi^2} y_t^4,
\label{quadr}\end{equation}
i.e. a factor of $\frac{f^2 y_t^2}{m_{\psi}^2}$ compared to conventional composite Higgs models.

In this paper, we UV complete the twin Higgs idea in the holographic framework \cite{MCHM_ADP,MCHM_ACP,MCHM_CNP,holographic_fermions} where the composite states are related to the KK tower of excitations in a RS setting (for other UV completions see \cite{Harnik-100TeV,SUSY_twin}). In this way, the Higgs potential is fully solvable - there are no logarithmic divergences and the dynamics is well defined up to the strong coupling scale. The spectrum consists of the SM particles, the mirror partners and KK excitations with various SM and mirror quantum numbers.

As in the original Twin Higgs \cite{Twin_Higgs}, our model requires an additional $Z_2$-breaking contribution to get the correct Higgs potential and to lift the mirror photon and the mirror partners of light states - in order to avoid potential constraints from cosmology \cite{Mirror_World,Planck}. We suggest a mechanism to softly break $Z_2$ in the strong sector, i.e in the bulk and on the IR brane. We apply this mechanism to generate the required $Z_2$ breaking holographically.

We assume that the $Z_2$ is an exact symmetry of the elementary sector, as in the original twin Higgs \cite{Twin_Higgs}. We do not seek a geometric origin for this discrete symmetry in this work. An alternative holographic approach is to forgo the exact $Z_2$ symmetry, but force the UV-brane boundary conditions to respect the exchange of $SM \to mirror$  (an example of this would be implementing the orbifold Higgs \cite{Orbifold_Higgs} in RS). This approach however, removes the $Z_2$ protection from the boundary gauge kinetic terms, and so mirror symmetry no longer protects the gauge contribution to the Higgs potential.

\section{The Model}
The model is set in a RS framework, with the UV and IR branes located at $z=L_0$ and at $z=L_1$. As will be evident from our choice of representations, the scales of the gauge and fermion excitations are the same: $m_{\rho}\approx m_{\psi} \approx m_{KK}=\frac{2}{L_1}$. In other composite Higgs models, the fermion excitations can be lighter than the gauge excitations to avoid tensions between naturalness \cite{tune-Panico,MCHM_ADP} and electroweak precision data (EWPD), not present in our model.

The bulk symmetry of the model is $SU(7)\times SO(8)$ corresponding to the global symmetry in the original twin Higgs model \cite{Twin_Higgs}, enlarged to accommodate an unbroken custodial symmetry \cite{RS_cust}. We choose SU(7) instead of the $SU(6) \times U(1)$ in \cite{Twin_Higgs} in order to avoid tree-level kinetic mixing between the neutral gauge bosons and their mirror partners. The bulk symmetry is broken on the IR brane into $SU(7) \times SO(7)$ with the Higgs as a PNGB in the $SO(8)/SO(7)$ coset. The symmetry on the UV brane is $(SU(3)\times SU(2)\times U(1))^{SM} \times (SU(3)\times SU(2)\times U(1))^{m}$.

The essence of the twin Higgs model is the $Z_2$ mirror symmetry, exchanging $SM \leftrightarrow mirror$. This symmetry is imposed on the UV-brane, i.e. as the symmetry of the elementary sector. It is embedded into the bulk symmetry $SU(7)\times SO(8)$ as the discrete subgroup exchanging the two $SO(4)$'s in $SO(8)$, and the two $SU(3)\times U(1)$'s in $SU(7)$. The mirror partners introduced by this $Z_2$ symmetry protect the Higgs potential from radiative corrections.

We choose the boundary conditions on the UV brane so that the conserved $U(1)$'s (hypercharge and mirror hypercharge) are generated by:
\begin{equation}
Y=T^3_R+\frac 4 3 T^7~,~Y^m=T^3_{mR}+\frac 4 3 T^7_m \label{hypercharge},
\end{equation}
where $T^3_{R}$ and $T^3_{mR}$ are the generators of $U(1)_R,U(1)_{R^m}\subset SO(8)$ and $T^7$ and $T^7_m$ are the generators of $U(1),U(1)^m \subset SU(7)$.

The Higgs is non-linearly realized in the vector representation $\bf 8_v$:
 \begin{equation}
 \Sigma=e^{-i \sqrt{2} T^a \frac{h^a}{f} }(0,0,0,0,0,0,0,1)^\intercal
 \end{equation}
where $T^a$ are 4 broken generators charged under $SU(2)^{EW}$. The other 3 broken generators are ``eaten" by the mirror gauge bosons.

In the quark sector the SM states $Q_L,t_R,b_R$ (and their mirror partners) are embedded in $\Psi_Q,\Psi_t,\Psi_b$ bulk multiplets - the $\bf 8_v,1,28$ representations of SO(8) and in the $\bf 7$ of $SU(7)$.

Under the $SU(2)^{SM} \times SU(2)^{m} \times U(1)^R \times U(1)^{R^m}$ subgroup of SO(8), $\bf 8_v$ and $\bf 28$ decompose as
\begin{eqnarray}
8_v&=&(2,1)_{\pm 1/2;0}+(1,2)_{0;\pm 1/2}\nonumber \\
28&=&(3,1)_{0;0}+(1,3)_{0;0}+(1,1)_{\pm1,0;0}+(1,1)_{0;\pm1,0}\nonumber\\
&+&(2,2)_{\pm\frac{1}{2},\pm\frac{1}{2}}.
\end{eqnarray}
$Q_L$, $t_R$ and $b_R$ are identified with the $(2,1)_{-\frac{1}{2};0}$, $(1,1)_{0;0}$ and $(1,1)_{-1;0}$ components. All of them are in the $(3,1)_{1/2,0}$ representation of $SU(3)^{SM}\times SU(3)^{m}\times U(1)_7 \times U(1)_7^m$. One can check in Eq.~\ref{hypercharge} that their SM hypercharge is reproduced.
Their mirror quarks, $Q^{m}_L$, $t^{m}_R$ and $b^{m}_R$ are the $(1,2)_{0;-\frac{1}{2}}$, $(1,1)_{0;0}$ and $(1,1)_{0;-1}$ components, and in the $(1,3)_{0,\frac{1}{2}}$ of $SU(3)^{SM}\times SU(3)^{m}\times U(1)_7 \times U(1)_7^m$.

In the 5d picture, only the SM fields and their mirror partners have Neumann b.c. on the UV brane. On the IR brane, $\Psi^Q$ is decomposed into the {\bf 1} and the {\bf 7} of SO(7). Both of these components have Neumann b.c. for the left handed chirality and we can write the IR mass term $m^1_q \Psi^{Q_1}_L \Psi^t_R$, where $\Psi^{Q_1}_L$ is the SO(7) singlet component of $\Psi^Q_L$.

The top sector holographic Lagrangian is given by:
\begin{equation}
L=\overline{\Psi^Q_L}\slashed p(\Pi^Q_0(p)+\Pi^Q_1(p)\Sigma\Sigma)\Psi^Q_L +\overline{\Psi^t}_R \slashed{p} \Psi^t_R+\overline{\Psi^Q_L} M_t(p) \Sigma \Psi^t_R
\end{equation}
where $M_Q(p),\Pi^Q_0(p)$ and $\Pi^Q_1(p)$ are calculated holographically \cite{holographic_fermions}. The top mass and its contribution to the Higgs potential are given by:
\begin{eqnarray}
  m_t&=&\frac{1}{2}\frac{v}{f}\frac{M_t(p \to 0)}{\sqrt{\Pi^Q_0(p\to 0)}} \nonumber\\
V(h) &=& -\frac{1}{8\pi^2f^2}2N_c \int p^3dp \left[\log\left(1+\frac{M_t^2(p)}{2p^2\Pi^Q_0(p)}\sin^2\frac{h}{f}\right)+ \right.\nonumber \\
&+& \left.2\log\left(1+\frac{\Pi^Q_1(p)}{2\Pi^Q_0(p)}\sin^2\frac{h}{f}\right)+ (\sin \leftrightarrow \cos) \right] \label{fullpot}
\end{eqnarray}
The ($\sin \leftrightarrow \cos$) part is the mirror partner contribution. The SU(2) gauge contribution can be calculated in a similar way \cite{MCHM_ACP}, while the contribution of the hypercharge boson is an order of magnitude smaller.
\begin{figure}[ht]
\includegraphics[scale=0.26]{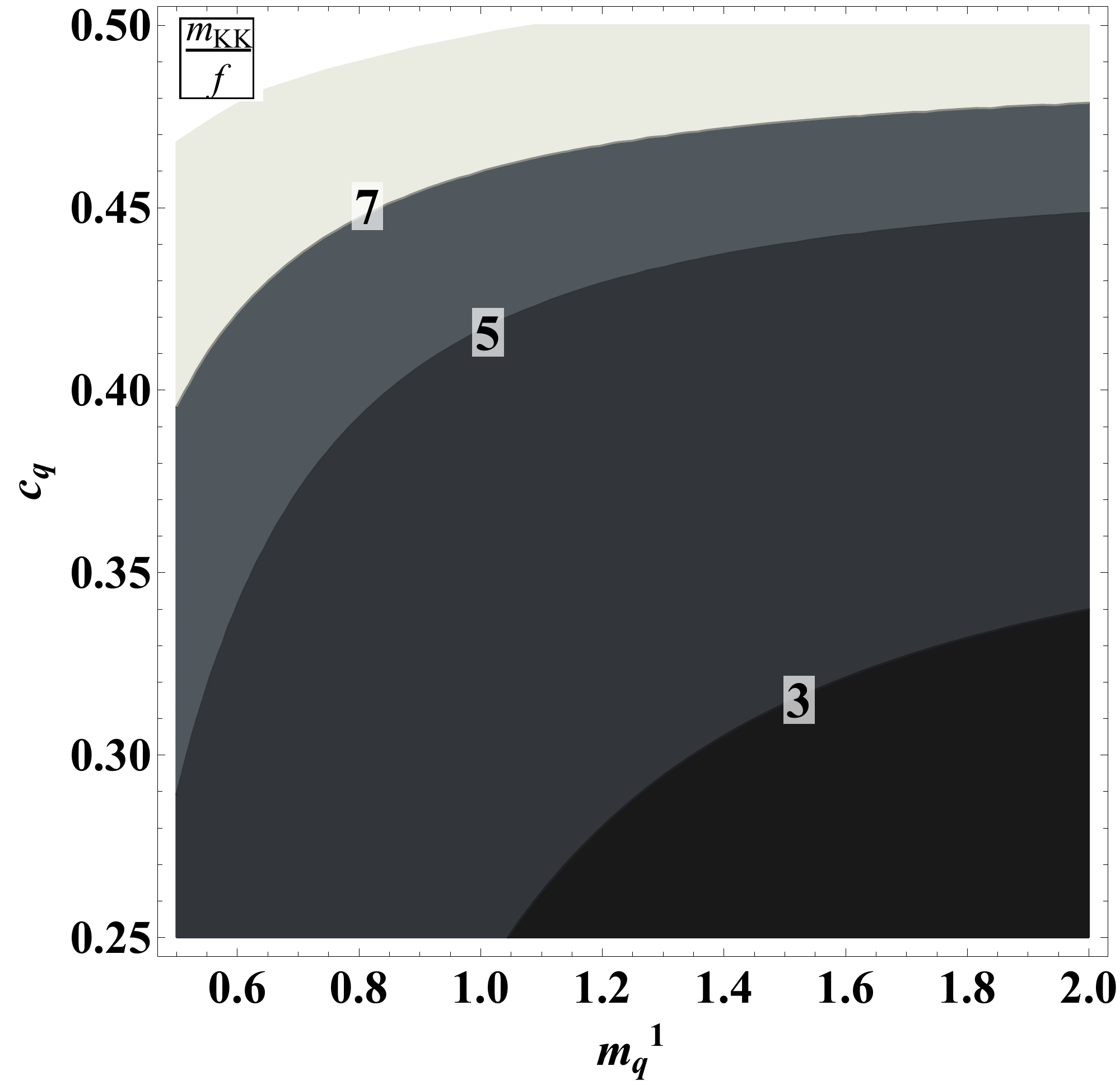}
\caption{\emph{The value of $\frac{m_{KK}}{f}$ required to reproduce the top mass, as a function of the bulk mass $c_q$ and the IR mass $m^1_q$ for an almost composite $t_R$, i.e. $c_u=\frac{1}{2}$.}
\label{mrho}}
\end{figure}

\begin{figure}[ht]
\includegraphics[scale=0.43]{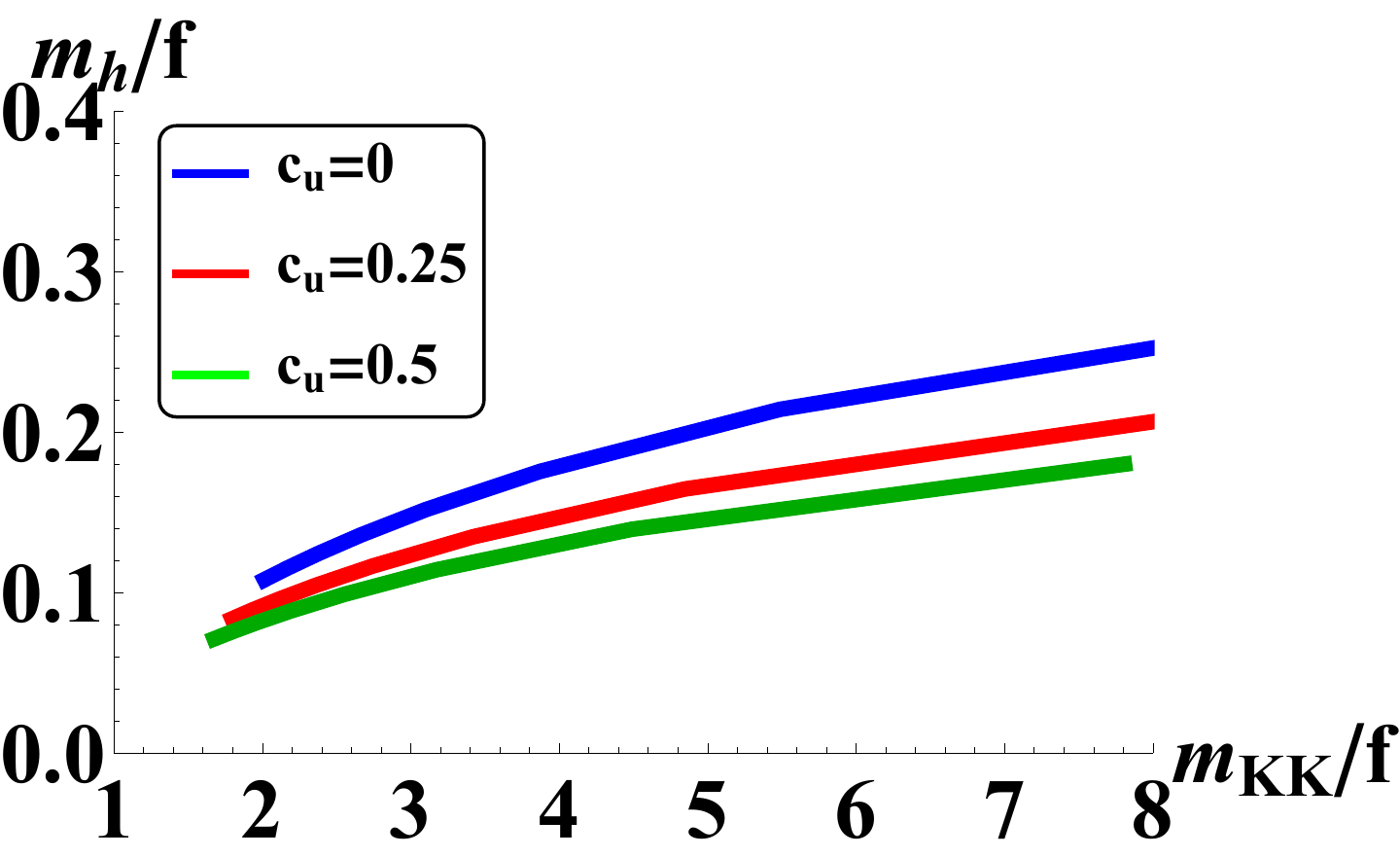}
\caption{\emph{The Higgs mass generated by the top and gauge sector. The weak dependence of the Higgs potential on $\frac{m_{KK}}{f}$ is a unique feature of the twin Higgs approach.}
\label{Potential}}
\end{figure}

The value of $\frac{m_{KK}}{f}$ required to reproduce the top mass is set by the bulk masses $c_q$ and $c_u$, and the IR mass $m^1_q$ (see Fig.~\ref{mrho}). In Fig.~\ref{Potential} we plot the typical Higgs mass generated by the top and gauge sector as a function of $\frac{m_{KK}}{f}$ for several values of $c_u$. In our choice of the parameter space $t_R$ is mostly composite, i.e. $c_u>0$. The resulting Higgs mass is typically $m_h\sim 0.2 f$, but the VEV is too high: $v \equiv f \sin\frac {<h>} f = \frac{1}{\sqrt{2}} f$.

As expected, $m_h$ is only weakly dependent on $m_{KK}$ because the quadratic divergence is cut off at the scale of the mirror partners rather than at the compositeness scale. For this reason, we avoid the generic tuning in composite Higgs models $\Delta>\left(\frac{m_{KK}}{400\text{ GeV}}\right)^2$ \cite{tune-Panico} and $m_{KK}$ can be naturally high. Nevertheless, an additional term is required to obtain a small $\frac{v}{f}$. This introduces a mild tuning, also present in CHM models \cite{tune-Panico} and the original twin Higgs model \cite{Twin_Higgs}. The additional term is:
\begin{equation}
V_s(h)=\mu^2_{s1} f^2\sin^2\frac{h}{f}-\mu^2_{s2}f^2 \sin^2\frac{h}{f}\cos^2\frac{h}{f} \label{singletpot}
\end{equation}

\begin{figure}[ht]
\includegraphics[scale=0.26]{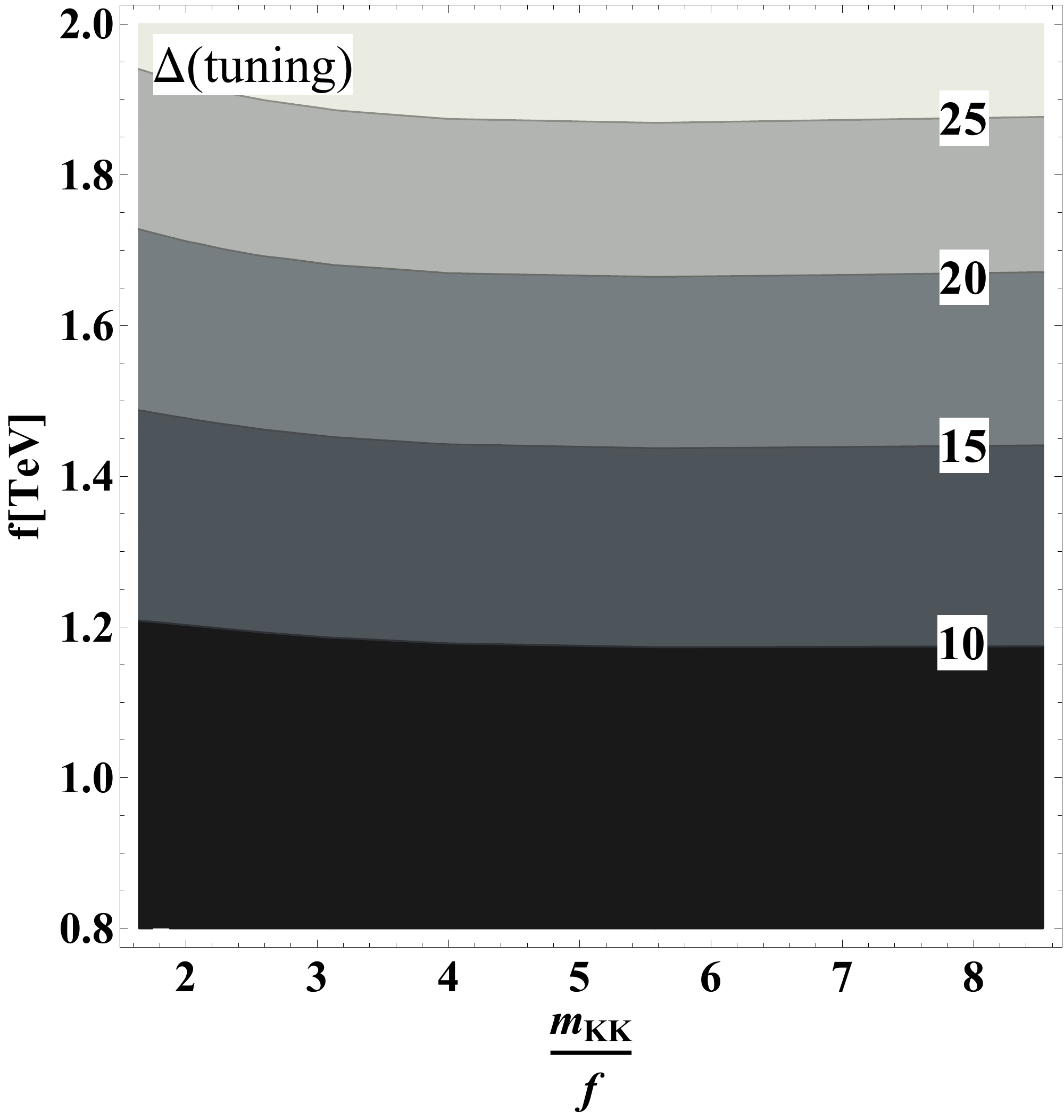}
\caption{\emph{The degree of tuning between the SM and the additional $V_s(h)$ contributions as a function of the pion scale $f$ and the ratio $\frac{m_{KK}}{f}$.}
\label{Tuning}}
\end{figure}
To understand the tuning in this model, it is useful to approximate the top and gauge contribution as $V(h)\approxeq -\alpha \sin^2\frac{h}{f}\cos^2\frac{h}{f}$. With this approximation we can calculate the vev and the tuning analytically:
\begin{equation}
\frac{v^2}{f^2} = \frac{\alpha+\mu^2_{s2}f^2-\mu^2_{s1}f^2 }{2\left(\alpha+\mu^2_{s2}f^2\right)}~,~ \Delta \approxeq \frac{f^2}{2v^2} \label{analytictuning}
\end{equation}
with the tuning defined as \cite{BG}:
\begin{equation}
\Delta=max(\frac{\partial \log m_Z}{\partial \log\mu_{s1}},\frac{\partial \log m_Z}{\partial \log\mu_{s2}}) \approx \frac{\partial \log m_Z}{\partial \log\mu_{s1}} \label{tune}
\end{equation}
While $\mu^2_{s1}$ is tuned to $\alpha+\mu^2_{s2}$, $\mu^2_{s2}$ is required to increase the generated quartic so that the mass of the Higgs is 125 GeV, especially for small $\frac{m_{KK}}{f}$. For large $f$ (and $m_{KK}$) the tuning is milder than in Eq.~\ref{analytictuning} due to an additional $\sin^4\frac{h}{f}$ term in the top contribution of Eq. \ref{fullpot} that scales as $\log\frac{m_{KK}}{v}$ \cite{MCHM_ACP,Twin_Higgs}. We plot the tuning calculated using the full potential of Eq.~\ref{fullpot} in Fig.~\ref{Tuning}.
We note that the $\mu_{s1}$ term is a $Z_2$ breaking term, akin to the one in \cite{Twin_Higgs} and the $\mu_{s2}$ term is $Z_2$ conserving.

To generate these terms we introduce a holographic $Z_2$ breaking mechanism. Such a mechanism can be used to keep the light mirror sector (the mirror partners of light states) independent of SM parameters. In this way, the model is potentially uncostrained by cosmological bounds, for instance the Planck limit on the effective number of neutrinos \cite{Planck}. Additionally, our model allows various DM scenarios within the light mirror sector. In the next section we describe the holographic $Z_2$ breaking mechanism and its use.

\section{A $Z_2$ breaking mechanism}

To include $Z_2$ breaking terms we first extend the bulk symmetry by an O(4), so that the full bulk symmetry is $SU(7)\times SO(8) \times O(4)$. The mirror $Z_2$ that previously acted within $SO(8)\times SU(7)$ now exchanges the two SU(2)s in the O(4) as well. This O(4) is spontaneously broken in the bulk to $SU(2)_4\times U(1)^m_4$, with only the $SU(2)_4$ unbroken on the IR brane. On the UV brane the hypercharge and mirror hypercharge are modified to:
\begin{equation}
Y=T^3_R+\frac 4 3 T^7+T^4~,~Y^m=T^3_{mR}+\frac 4 3 T^7_m+T^4_m\label{hyperZ2}
\end{equation}
The mirror photon is now massive, due to the breaking of $U(1)^m_4$ on the IR brane. The PNGBs from the $O(4)$ breaking are SM singlets.

The breaking in the bulk is translated to different bulk masses for different components of O(4) multiplets. As a result, $Z_2$ partners in these multiplets are localized differently in the bulk and have different Yukawa couplings.

We embed the leptons and first two quark generations in the $\bf 6$ of O(4). They are identified with the $T^4_m=T^4=0$ components of the $SU(2)^m_4\subset O(4)$ triplet within this multiplet. Accordingly, their mirror partners are the $T^4_m=T^4=0$ components of the $SU(2)_4 \subset O(4)$ triplet. The masses of the mirror partners of the light states are then arbitrary in our model.

We now turn to produce the terms from Eq.~\ref{singletpot}. The $Z_2$ breaking term is generated by a SM singlet embedded in the $\bf 28$ of SO(8) and in the $\bf 6$ of O(4). The singlet is the neutral component of the $SU(2)_R^m \subset SO(8)$ triplet and the $SU(2)^m_4\subset O(4)$ triplet. Its mirror partner is also a SM singlet with a different bulk mass due to the $O(4)$ breaking. We assume it to be localized sufficiently far from the IR brane so that it doesn't contribute to the Higgs potential. To create the $Z_2$ conserving term we further introduce a SM singlet fermion embedded in the $\bf 35_v$ of SO(8), which is its own mirror partner. The two new multiplets couple on the IR brane.

On the UV brane only the SM singlet components of $\bf 28,35_v$ have Neumann b.c. for the left handed chirality. The IR brane the b.c. are:
\begin{eqnarray}
\Psi^{21}_L(+)~,~\Psi^{7}_L(+)~ &\in&~ \Psi^{28}_L \nonumber\\
\Psi^{27}_R(+)~,~\Psi^{7}_R(+)~,~\Psi^{1}_R(+) ~&\in&~ \Psi^{35_v}_R
\label{IRbc}\end{eqnarray}
with an IR mass term $m_7\bar {\Psi}^{7}_L \Psi_R^7$.
The Higgs potential generated by the new fermions is
\begin{eqnarray}
V(H)&=&\mu_{s1}^2(c_{28},c_{35v},m_7) f^2 \sin^2\frac{h}{f} \nonumber\\
&-&\mu^2_{s2}(c_{28},c_{35v},m_7)f^2\sin^2\frac{h}{f}\cos^2\frac{h}{f}
\end{eqnarray}
The free parameters in this case are the bulk masses and the IR-brane mass, denoted by $c_{28},c_{35v}$ and $m_7$. They are selected to reproduce values of $\mu_{s1},\mu_{s2}$ in the relevant range (See Eq.~\ref{analytictuning}). The tuning is now given by
\begin{equation}
\Delta=\frac{\partial \log m_Z}{\partial \log \mu_{s1}} \cdot max\left(\frac{\log \mu_{s1}}{\log \left\{c_i,m_7 \right\}}\right)
\end{equation}
with typically $\frac{\partial \log \mu_{s1}}{\partial \log \left(c_i,m_i\right)}<1$ for $f\sim 1$ TeV in the desired area of the parameter space ($c_{28}> 0$, $c_{35s}<0$, $m_7\sim 1$). The new singlets are massless, but can be easily lifted with no consequence to the Higgs potential.

\begin{table}
\begin{tabular}{*4c}
\toprule
\emph{Fermion} &  $SU(7)$ & $SO(8)$ &$~SU(2)_4\times U(1)^m_4 \subset O(4)$ \\ \colrule
\multicolumn{4}{l}{\bf Quarks: Third Generation  } \\
$\Psi_L^Q$ & {\bf 7}& {\bf 8} & {\bf 1} \\
$\Psi_R^t$ & {\bf 7}& {\bf 1} & {\bf 1} \\
$\Psi_R^b$ & {\bf 7}& {\bf 28} & {\bf 1} \\
\multicolumn{4}{l}{\bf Light Quarks and Leptons } \\
$\Psi_L^{Q/L}$ & {\bf 7/1}& {\bf 8} & $({\bf 1},0)\in {\bf 6}$   \\
$\Psi_R^{u/\nu}$ & {\bf 7/1}& {\bf 28} or {\bf 1} & $({\bf 1},0)\in {\bf 6} $\\
$\Psi_R^{d/e}$ & {\bf 7/1}& {\bf 28} & $({\bf 1},0)\in {\bf 6}$ \\
\multicolumn{4}{l}{\bf New Singlets}\\
$\Psi^{35}_R$ &{\bf 1} &{\bf 35v} &  ${\bf 1}$\\
$\Psi^{28}_L$ &{\bf 1} &{\bf 28} & $({\bf 1},0)\in {\bf 6}$ \\
\botrule
\end{tabular}
\caption{The fermion content of the model with their representations under the bulk gauge group $SU(7)\times SO(8) \times U(1)_4
\times U(1)^m_4$ \label{tablefermion}.}
\end{table}

The fermion content of the model is given in Table \ref{tablefermion}.

\section{Phenomenology}

Generally in composite Higgs models, bounds on $m_{KK}$ imply lower limits on the amount of tuning in the Higgs potential. The most relevant bounds are from VLQ searches \cite{VLQ_searches,VLQ_whitepaper}, that probe the top excitations directly. The mass of the gauge excitations is constrained by EWPD \cite{ew-MCHM,tune-Panico}, but is only loosely related to the tuning. In particular, for $m_{KK}$ larger than the LHC reach of roughly 2 TeV \cite{VLQ_whitepaper}, the tuning is at least at the percent level \cite{tune-Panico}.

In our model, the tuning is almost independent of $m_{KK}$ and depends primarily on $f$ (see Fig.~\ref{Tuning}). This allows us to choose $m_{KK}$ as high as calculability allows us: $\frac{m_{KK}}{f} < 4 \pi$. We assume that $m_{KK}\lsim 7$ TeV and $f\sim 1$ TeV, so that tuning is ${\cal O}(10\%)$.

The EW constraints \cite{ew-MCHM,tune-Panico} are easily satisfied due to the high scale of excitations ($m_\rho\approx m_{KK}>3$ TeV) and due to the custodial symmetry of the bulk and IR brane \cite{MCHM_ADP}.


The fermion KK excitations lie above $m_{KK}$ and include a $2_{7/6}$ VLQ with $m_{7/6} \lsim 7$ TeV, and a $2_{1/6}$ VLQ with $m_{1/6}\lsim 10$ TeV. These states are naturally  beyond the LHC reach, but are potentially accessible in a future 100 TeV collider (see \cite{VLQ-100TeV}).

Finally, precision Higgs measurements can produce bounds on $f$, as in any PNGB Higgs model, due to the modification of all the partial widths by a $1-\frac{v^2}{f^2}$ factor. While the LHC can probe $f$ up to $900$ GeV \cite{Composite}, future leptonic colliders can produce significantly higher bounds on $f$ \cite{ILC_HIGGS_WIDTH}. Additionally, the Higgs can now decay to mirror quarks, predominately to the mirror-bottom, whose Yukawa is $y_{b^m}=\frac{f}{v}y_{b}$. The invisible width  $\Gamma_{inv}\approx 0.5 \frac{v^2}{f^2}\cdot\Gamma^{SM}_{bb}$ can be probed at future leptonic colliders \cite{ILC_HIGGS_WIDTH}.

\begin{figure}[ht]
\includegraphics[scale=0.37]{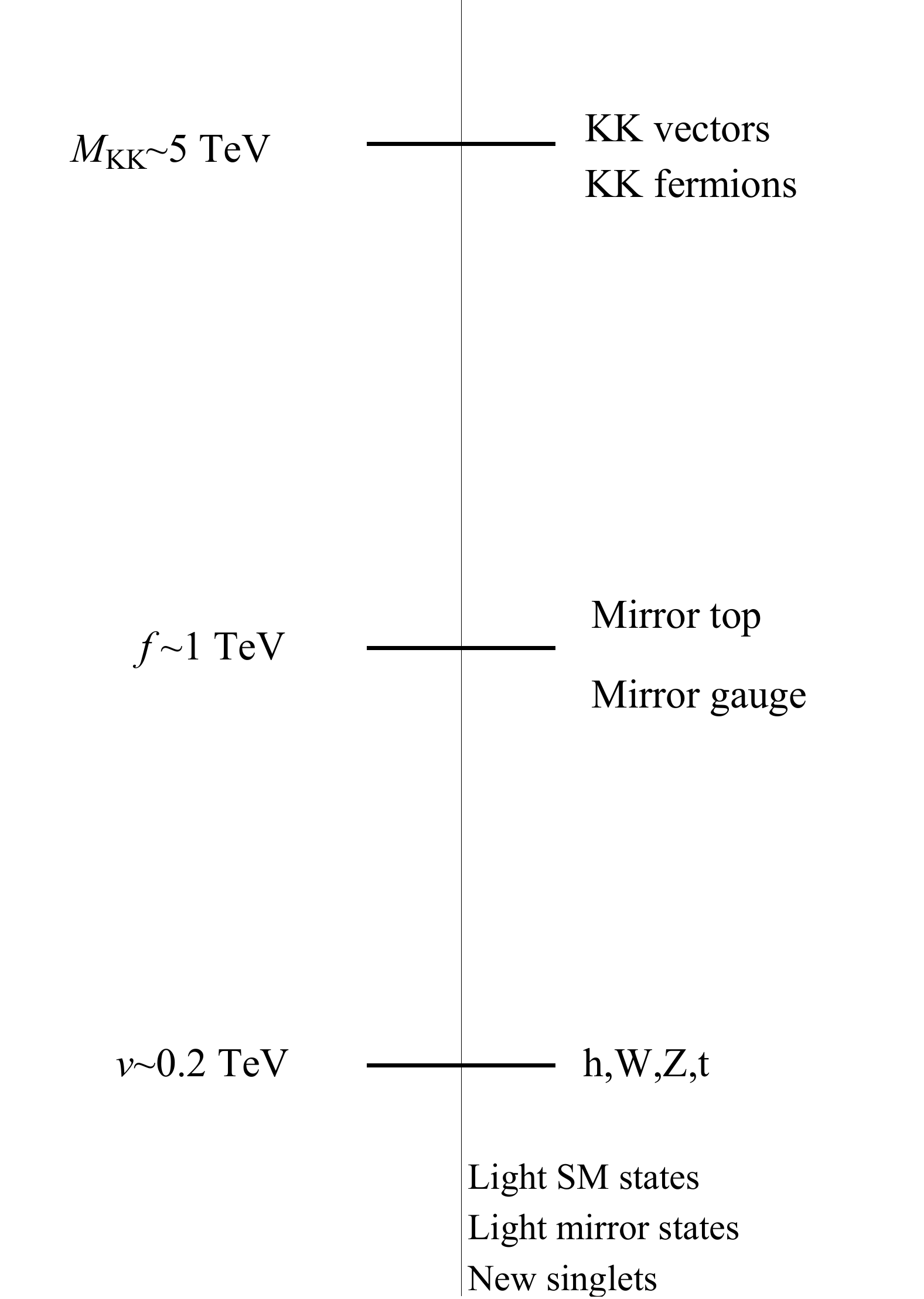}
\caption{\emph{The spectrum of the Holographic Twin Higgs}.}
\end{figure}

\section{Summary and Conclusions}

In this work we presented the holographic twin Higgs. This model is an implementation of the twin Higgs idea \cite{Twin_Higgs} within the framework of 5d AdS background. The bulk symmetry group is $SU(7)\times SO(8)$, broken on the IR brane into $SU(7)\times SO(7)$ and on the UV brane into $(SU(3)\times SU(2)\times U(1))^{SM} \times (SU(3)\times SU(2)\times U(1))^{m}\times Z_2$. The $Z_2$ symmetry on the UV brane is identified with the bulk symmetry operator exchanging $mirror \leftrightarrow SM$. Additionally, we introduced a simple mechanism that breaks the $Z_2$ holographically using an O(4) extension of the bulk symmetry.

The contribution to the Higgs potential, generated via SM fermion and gauge loops, is cut off by the mass of the mirror partners rather than the KK scale, as in the conventional CHM. As a result, values of $m_{KK}$ beyond the reach of the LHC are natural. However, an additional $Z_2$ breaking contribution is required to get $v<f$. This contribution is ${\cal O}(10\%)$ tuned to get a reasonable $\frac{v}{f}\sim \frac{1}{4}$. Our $Z_2$ breaking mechanism is utilized to generate it holographically, as well as to make the light mirror sector independent of the SM.

The particle spectrum in our model is:
\begin{enumerate}
\item Top and gauge mirror partners: SM singlets with ${\cal O}\left(TeV\right)$ masses.
\item Light mirror states and new singlets: SM singlets, possibly DM candidates with arbitrary masses below the EW scale. May be probed as invisible Higgs width at future colliders.
\item KK excitations: Vector-like quarks and heavy gauge bosons, with ${\cal O}(5\text{ TeV})$ masses, beyond the reach of the LHC. May be probed at a 100 TeV collider.
\end{enumerate}

\section{Acknowledgements}

We are extremely grateful to Kaustubh Agashe for many valuable discussions. We also benefited from discussions with Luca Vecchi, Yael Shadmi, Amarjit Soni, Shaouly Bar-Shalom, Andrey Katz, Yevgeny Katz, Gilad Perez, Brian Batell, Oren Bergman. The authors acknowledge research support from the Technion.

\end{document}